# SYNTHETIC OIL GELS WITH ORGANOCLAYS IN THE FORMULATION OF MAGNETORHEOLOGICAL FLUIDS

## JOSÉ H. R. ROCHA*, JÚLIO G. F. MANUEL,* ANTONIO J. F. BOMBARD*

\* Universidade Federal de Itajubá – UNIFEI, Instituto de Física e Química
Av. BPS 1303, Itajubá – MG, 37.500-903, Brazil
e-mail: bombard@unifei.edu.br

**Abstract**. Magnetorheological fluids (MRF) are smart composite materials that, under an external magnetic field, show a reversible solid-liquid transition in less than 10 ms. This study aimed to evaluate which organoclays would jellify a synthetic oil for the formulation of MRF. Three dispersant additives for carbonyl iron powder were evaluated. Fifteen different gelling additives from four clay families, bentonites, hectorites, montmorillonites, and mixed mineral thixotropes (MMT), were dispersed in oil only, keeping the same concentration, without iron particles. The gels were then tested through amplitude and frequency sweeps in oscillatory rheometry to evaluate their viscoelastic behavior. The thixotropy of the gels was measured through the "three-interval" test in a rheometer. After selecting the best gelling additive to prepare the MRF, three dispersing additives had their rheology evaluated to determine the best magnetorheological effect and redispersibility after one year of sample preparation.

In the linear viscoelastic region, all MMT clays resulted in a weak viscoelastic gel (G' ~ 100 to 300 Pa and G" ~ 30 to 50 Pa). Some of the bentonite clays jellified, and others did not. The best organoclays were montmorillonites and hectorites, which formed consistent viscoelastic gels (G' ~ 1 to 5 kPa and G" ~ 70 to 250 Pa).

The best organoclay presented a yield stress $\sigma_0$ = (42 ± 3) Pa, a storage modulus G' = (2690 ± 201) Pa, and a cohesive energy density (CED) = 98 mJ/m³, and it was selected to explore the rheology of MRF with three dispersant additives: octan-1-ol, octan-1-amine and L-α-Phosphatidylcholine. All the MRFs were prepared using carbonyl iron powder HS (BASF SE) in oil gels and with the same organoclay. All three dispersant additives showed a thixotropic recovery above 100% in the three-interval test.

Regarding the redispersibility after one year, the MRF formulations with octan-1-amine and lectithin were reproved, as they reached normal force peaks of 19 and 24 N, while the work was 28 and 415 mJ, respectively. The best MRF was formulated with octan-1-ol, and resulted in a normal force of 0.33 N and 3.4 mJ at 35 mm of vane penetration. Therefore, we conclude that the MRF with octan-1-ol and montmorillonite #6 showed a better balance between thixotropy, MR effect, and, above all, good redispersibility.

# 1 INTRODUCTION

Although the term "composite" originated from the combination of polymeric resins with reinforcing fibers [1], magnetorheological (MR) materials (whether MR fluids, MR Gels, or MR Elastomers, etc.) are also composite materials [2]. In the case of magnetorheological fluids (MRF), the matrix is a continuous liquid phase, and the solid magnetic particles are the dispersed phase [3]. For magnetorheological elastomers (MRE), the continuous matrix is an elastomer, such as silicone rubber [4]. For magnetorheological gels (MRG), it can either be a gel formed by a copolymer dissolved in oil [5], or an inorganic gel such as clays (natural, if the liquid phase is aqueous, or chemically modified, if the medium is a solvent or a synthetic oil derived from hydrocarbons), for example [6].

In the case of MRF, the simplest formulations consist only of the dispersed magnetic phase and the continuous liquid medium. The most commonly used material during preparation is carbonyl iron powder (CIP). This material has a purity above 97% and consists of zero-valence metallic iron. As a continuous phase, the most common liquids used to prepare MRF are water or aqueous solutions, oils (mineral, silicone, synthetic, vegetable, etc.), perfluoropolyethers, glycols, ionic liquids, among a myriad of other MRF formulations [7]. Although some exotic liquid metals such as Gallium, Mercury, and Galinstan have been reported in the literature, these materials can form metallic alloys with the CIP, resulting in amalgams [8]. They are, therefore, not exactly MRF. Water-based MRF are commonly applied in MR polishing, while in the automotive industry, MR devices (shock absorbers, brakes, clutches, etc.) usually have oils as the liquid phase (generally poly-alpha-olefin oil or silicone oil) [7].

While physicists usually avoid additives that would complicate the analysis of their experimental findings (and the comparison with their theoretical predictions), engineers and scientists who design MR devices (and those interested in preparing MRFs) cannot avoid formulating MRFs with additives. In this context, the most used additives are dispersants, gelling agents, lubricants, anti-wear, anti-corrosives, anti-foams, etc. [7].

An inherent and inevitable problem for anyone who has ever formulated an MRF is the sedimentation of the dense iron particles in a liquid medium with a much lower density. Another relevant problem, although this one is easier to solve, is the coagulation/aggregation of the particles caused by the van der Waals forces and, even more, the residual remanence of the CIP particles [9]. In this case, correctly selecting the dispersant additives can solve the problem. As for the former problem, the best one can achieve is to delay sedimentation and ensure an easy redispersion. However, completely preventing sedimentation is impossible. Some might argue that with the advent of MRF's "younger brothers", such as MR Elastomers, MR Gels, or MR Foams, the sedimentation problem has been solved and eliminated. Nevertheless, there is a significant reduction in the range of control of these materials compared to conventional MR fluids. MR elastomers are prepared by dispersing magnetic particles in a rubber matrix, which can be cured under an applied magnetic field, pre-orienting the particles in the typical chain structures of MR materials, or even without a magnetic field, and with this the particles generally are left randomly in the rubber matrix. But the uses and applications of MRE are very different from those of MRF. MRE has only found applications where elasticity control is important.

It is well established in patents and scientific literature on MRF that pure carrier fluids, without gelling additives, suffer from rapid sedimentation and coagulation of dense iron particles. Hence the use of gelling additives, such as organomodified clays. A thixotropic gel suitable for MRF formulations must exhibit shear thinning and thixotropic behavior, and the higher the storage modulus G' values in the LVE region and the yield stress of the gel, the more effective the additive is in gelling the carrier fluid. Besides, synthetic oil in a 'gel' state, when analyzed by oscillatory rheometry, must present two main characteristics: a) the value of G' >> G'' (the loss modulus), in several orders of angular frequency $\omega$, and b) a value of G' almost constant, virtually invariable with the frequency.

Among the most common gelling additives cited in the literature and in patents of MRF, fumed silicas and clays (natural or organomodified) stand out [10, 11]. This work aimed to investigate, among four main types of clay, which one is the most suitable, out of 15 evaluated clays, to gel a synthetic polyalphaolefin (PAO) oil. These classes were: montmorillonite, bentonite, hectorite, and mixed mineral thixotrope (MMT) [12]. In a subsequent step, we chose a single clay (the one that was most effective in gelling the PAO oil, as measured through a higher G' value) and evaluated among three dispersant additives: 1) octan-1-ol [13,14], 2) octan-1-amine [15,16] and 3) L-$\alpha$-Phosphatidylcholine (soybean lecithin) [17,18], which showed the best performance. The following responses were analyzed: a) the MR effect under a magnetic field (on-state), b) the thixotropic recovery without a magnetic field (off-state), c) the redispersibility of the samples at rest, under normal gravity, and without an applied field, in the periods of 3, 7, 30 and 365 days.

## 2 EXPERIMENTAL

### 2.1 Preparation of the thixotropic gels

To prepare the gels, fifteen types of clay, organized in four groups, were dispersed in polyalphaolefin oil (Durasyn PAO 162, INEOS), totaling 5 wt.% of the oil's mass. The gels were then homogenized with an IKA Ultra Turrax T-18 mechanical disperser at 6,000 rpm for 30 seconds, followed by the addition of the chemical activator: propylene carbonate - water 95:5, and 33.3 wt.% of the clay's mass [19]. The mixture was homogenized again at 20,000 rpm for a minute, and their rheometry was investigated.

### 2.2 Rheometry of the gels

The samples that were jellified were taken to a rheometer (Anton Paar Physica MCR 301) for an oscillatory sweep test to measure the yield stress $\sigma_0$ and the storage modulus G'. The organoclay which was the most effective in jellifying the PAO oil, and resulted in a gel sample with greater values of G' and $\sigma_0$ was selected as a thixotropic additive to prepare the MRF. An oscillatory frequency sweep was also performed to analyze the complex viscosity $\eta*$ and to compare the storage and loss moduli (G' and G'').

### 2.3 Preparation of the magnetorheological fluids

After carefully selecting the best additive, three MRFs were prepared with 50 vol% of carbonyl iron particles (d = 7.86 g.cm$^{-3}$), 49 vol% of carrier liquid (PAO + activator + dispersant), and 1 vol% of clay. A different dispersant was used in each preparation, and they were labeled: a) octan-1-ol, b) octan-1-amine, and c) lecithin. Firstly, the polyalphaolefin oil was weighed, and the dispersant was added (0.7 wt% of the carbonyl iron particles). Then, the iron particles were added, and the mixture was homogenized with an IKA Ultra Turrax T-18 mechanical disperser at 6,000 rpm for 2-3 minutes. The clay was added, the mixture was homogenized again, and the activator was added [19]. Finally, the sample was stirred vigorously at 20,000 rpm for 60 seconds to jellify the clays, and their rheology was studied. Table 1 shows the composition with the respective percentages by mass and volume of each of these 3 MRF formulations.

**Table 1**: Composition (in weight/weight % and volume/volume %) of the three MRFs evaluated.

| Material | MRF 1 Octan-1-ol | | MRF 2 Octan-1-amine | | MRF 3 Lecithin | |
|---|---|---|---|---|---|---|
| | Wt. % | Vol. % | Wt. % | Vol. % | Wt. % | Vol. % |
| CIP HS | 90.82 | 50.0 | 90.61 | 50.0 | 90.80 | 50.0 |
| PAO oil | 7.81 | 45.3 | 8.08 | 44.5 | 7.85 | 45.4 |
| Dispersant | 0.71 | 3.1 | 0.71 | 3.9 | 0.70 | 3.0 |
| Organoclay | 0.47 | 1.0 | 0.45 | 1.1 | 0.47 | 1.0 |
| Clay Activator | 0.19 | 0.6 | 0.15 | 0.5 | 0.18 | 0.6 |

### 2.4 Magnetorheology

After preparation of the MRF, the following tests were performed:

*2.4.1 Amplitude sweep (off-state)*: To obtain the linear viscoelastic range, the prepared MRF were subjected to amplitude sweeps in the absence of an applied magnetic field at 25 ºC. A cone-and-plate configuration was used (diameter of 25 mm) with a truncation d = 0.056 mm and frequency ω = 10 rad/s. The strain γ ranged from 0.001% to 100% [20].

*2.4.2 Frequency sweep (off-state)*: To analyze the complex viscosity and to obtain the values of G' and G'', frequency sweeps were carried out in the absence of an applied magnetic field at 25 ºC. A cone-and-plate configuration was used (diameter of 25 mm), with a truncation d = 0.056 mm and constant strain γ = 0.05% (linear viscoelastic range). The angular frequency ω was swept in decreasing order, from 500 rad/s to 0.05 rad/s [21].

*2.4.3 Amplitude sweep (on-state)*: The MRF were subjected to amplitude sweeps under an applied magnetic field at 25 ºC. A plate plate configuration was used (diameter of 20 mm), with a gap d = 0.5 mm and frequency ω = 10 rad/s. The strain γ ranged from 0.001% to 100%, and three conditions of current and magnetic flux density were analyzed: 1) $I_1$ = 0 A, then $B_1$ = 0 mT, 2) $I_2$ = 1 A, so $B_2$ = 192 mT, and 3) $I_3$ = 2 A, with $B_3$ = 370 mT [20].

*2.4.4 Magneto sweep*: The MRF were then subjected to a mag sweep, and the magnitude of the applied magnetic field was measured with a gaussmeter at 25 ºC. A plate plate configuration was used (diameter of 20 mm), with a gap d = 0.5 mm and constant frequency ω = 10 rad/s. The strain γ ranged from 0.01% to 100%, and the current linearly ramped from 0 – 3 A [22].

*2.4.5 Thixotropy*: The thixotropy of each MRF was analyzed without a magnetic field (off-state), with a parallel plate configuration (20 mm of diameter) and a gap d = 1 mm. The measurements were carried out through the so-called "step test with three intervals": in the first and third intervals, the sample was sheared at a shear rate $\dot{\gamma} = 0.01$ s$^{-1}$. In the second interval, the sample was sheared at $\dot{\gamma} = 100$ s$^{-1}$. The shear viscosity was computed during the entire period. To readers interested in measuring the thixotropy of rheological materials in detail, a reference from Mezger is recommended [23].

*2.4.6 Redispersibility*: Finally, each sample had its redispersibility measured with the rheometer after sedimentation at rest and under normal gravity. The normal force and the work done were then computed. The tests were repeated after three days, one week, one month, and one year [24].

## 3 RESULTS AND DISCUSSION

### 3.1 Organoclay gels and amplitude sweep

Figure 1 shows the storage modulus as a function of the strain amplitude for four OM clays: #6 is a smectite (or montmorillonite), #9 is MMT, #13 is a hectorite, and #15, is a bentonite. From Figure 1, one can see that the MMT generates the weakest gel, with only G' ≈ 200 Pa in the linear viscoelastic (LVE) region, while the other clays gave stronger gels, increasing in the following order: bentonite (G' ≈ 750 Pa) < hectorite (G' ≈ 2,800 Pa) < smectite (G' ≈ 4,600 Pa). With these results, the best OM clay #6 (smectite) was chosen, and all MRFs were prepared with this clay.

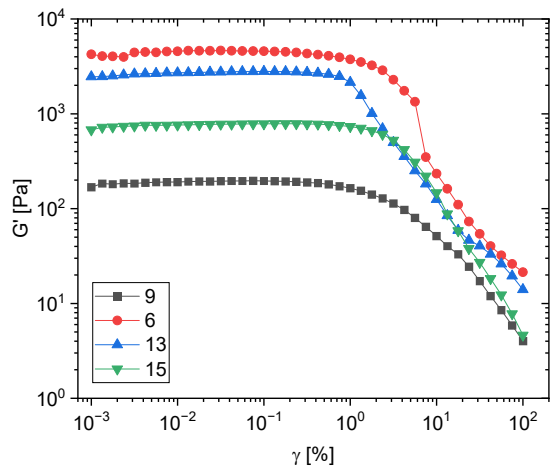

**Figure 1**: The storage modulus G' as a function of the amplitude strain γ for the gels obtained with OM clays

and PAO oil only, at 25 °C.

According to Ross-Murphy, if the complex viscoelastic modulus G* is divided by $G^*_{\gamma \to 0}$ (i.e., the G* value when the strain γ tends to zero), and a graph of this ratio is constructed as a function of the strain γ (%), one can classify the material as a 'weak gel' if the ratio $G^*/G^*_{\gamma \to 0}$ drops almost to zero, when the strain < 1%. If this ratio remains constant around one for strain values up to ~100%, the material can be considered a strong gel [25]. In this case, one can see that the gels formed with different clays resulted in weak to moderate gels. On the other hand, according to Mezger, whenever G' > G" in an oscillatory amplitude sweep, the value of G' can be considered a measure of the "gel strength" [26]. Therefore, the gelling agents based on smectite (#6), and hectorite (#13) clays resulted in stronger gels than bentonite (#15) or MMT (#9) clays.

## 3.2 Organoclay gels and frequency sweep

Figure 2 shows the viscoelastic moduli plotted as a function of the frequency sweep (under constant amplitude strain in the LVE region) to verify the behavior of gel (#6): a strong gel, a weak gel, or a viscoelastic liquid. In this figure, one can see that the storage modulus G' was around 2 kPa, while the loss modulus G" was around 200 Pa in the entire angular frequency range. The complex viscosity η* continuously decreases with increasing frequency, as is typical in materials with shear thinning behavior. Besides, the complex viscosity has a constant slope of approximately −1, which is typical of a gel. According to Steffe [27] and Almdal *et al.* [28], a 'gel', when analyzed by oscillatory rheometry, must present two main characteristics: a) a value of G' greater than G" in several orders of frequency and b) a constant G', virtually invariable with the frequency. Therefore, as G' ≈ ten times greater than G", and the two modules are relatively constant and independent of frequency, Figure 2 corroborates the rheological behavior of a gel.

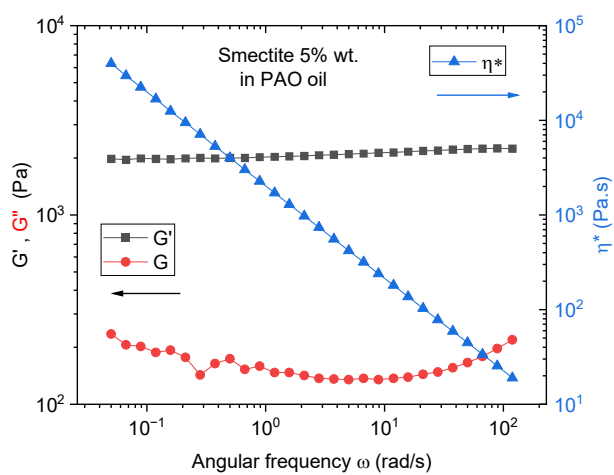

**Figure 2**: Viscoelastic moduli G' and G" (left axis) and complex viscosity η* (right axis) as a function of the angular frequency ω for the strongest gel prepared with OM smectites (5 wt%) in PAO oil only, at 25 °C.

### 3.3 The yield stress of the organoclay gels, by oscillatory amplitude sweep

Figure 3 shows the storage modulus as a function of the shear stress for gel (#6). In this type of plot, the value where the storage modulus G' decreases abruptly corresponds to the yield stress σ itself. The measurements were performed in triplicate, with previously untested samples.

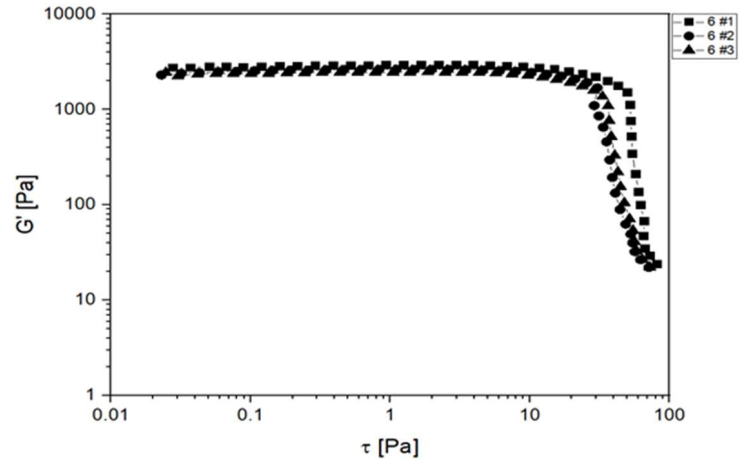

**Figure 3**: The storage modulus as a function of the shear stress for the best gel formulation, an OM smectite (5 wt%) in PAO oil only, at 25 °C.

From figure 3, it can be seen that the average value of the yield stress, in which the storage modulus G' has a sharp drop, is around 40 Pa.

### 3.4 The amplitude sweep and the cohesive energy density

With the oscillatory amplitude sweep curves, it is also possible to compute an important parameter for the gel, the so-called cohesive energy density (CED). It can be calculated with the help of Equation 1:

$$C.E.D. = \frac{1}{2} G'(\gamma_{cr})^2 \tag{1}$$

where G' is the storage modulus value at its plateau region (LVE range) and $\gamma_{cr}$ is the critical strain (without units and not in percentage), that is, the maximum value of strain before the storage modulus G' begins to decrease [29]. Figure 4 shows the procedure used to compute the cohesive energy density of gel (#6):

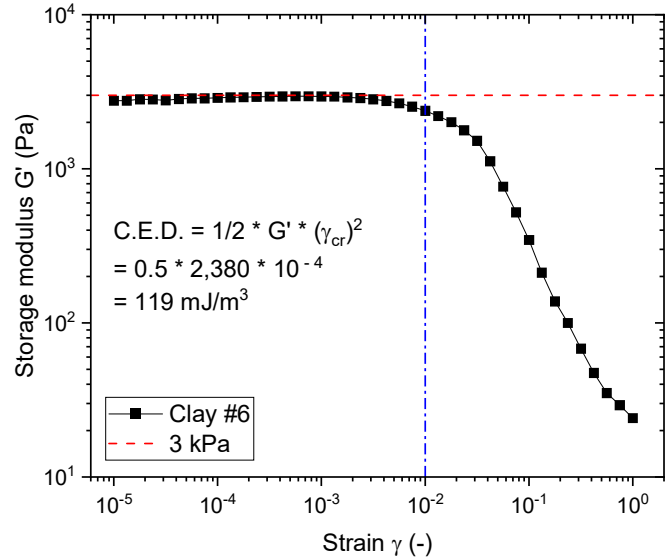

**Figure 4**: The storage modulus as a function of strain for the best gel formulation, an OM smectite (5 wt%) in PAO oil only, at 25 °C. Although the dashed lines are arbitrary, the region where there is a 20% reduction of the maximum value of G' can be taken as the critical strain. In this case, G' = 2,380 Pa, $\gamma_{cr}$ = 0.01 (no units) and, therefore, the Cohesive Energy Density ≈ 120 mJ/m³.

The cohesive energy density is a useful parameter to evaluate the strength of inorganic and organic gels, as it measures the mechanical energy required to break weak interactions. The greater its value, the more resistant a material is to flow and slower must be the sedimentation of iron micro particles through the gel [29].

## 3.5 Amplitude sweep (on-state):

After choosing clay #6 as an additive, the MRF were prepared, and their rheology was investigated. Firstly, the yield stress values σ of each fluid were computed from a Storage Modulus versus Shear Stress plot, as shown in Figure 5.

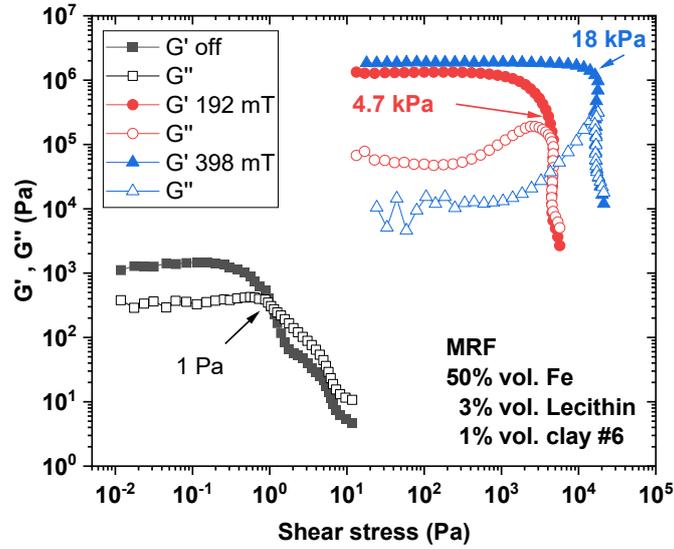

**Figure 5**: Amplitude sweep test for the MRF prepared with clay #6 and lecithin as the dispersant, without a magnetic field (black squares), and under B = 192 mT (red circles), or B = 398 mT (blue triangles). Filled symbols: G'. Empty symbols: G". The arrows indicate the yield stress values used in the calculation of the relative MR effect.

The relative magnetorheological effect of each fluid was calculated according to Equation 2:

$$MR_{relative} = \frac{(\sigma_B - \sigma_{off})}{\sigma_{off}} \qquad (2)$$

where $\sigma_B$ is the yield stress value on the on-state, and $\sigma_{off}$ is the yield stress on the off-state.

### 3.6 Thixotropy

As for the thixotropic response, all MRF formulations showed fast (less than 2 seconds) and complete (above 100%) thixotropic recovery. Therefore, we chose not to present the corresponding graph because, qualitatively, all additives were satisfactory regarding this property, which also demonstrates that it depends almost exclusively on the chosen clay.

### 3.7 Redispersibility

Finally, the redispersibility of each fluid was analyzed, and the results are shown in Figure 6. For brevity and to avoid showing qualitatively unimportant results, we will only present the redispersibility results after one year. As an illustration, figure 7 shows a picture of the test tubes, containing part of the MRF formulated and tested in the redispersibility, after 4 years at rest.

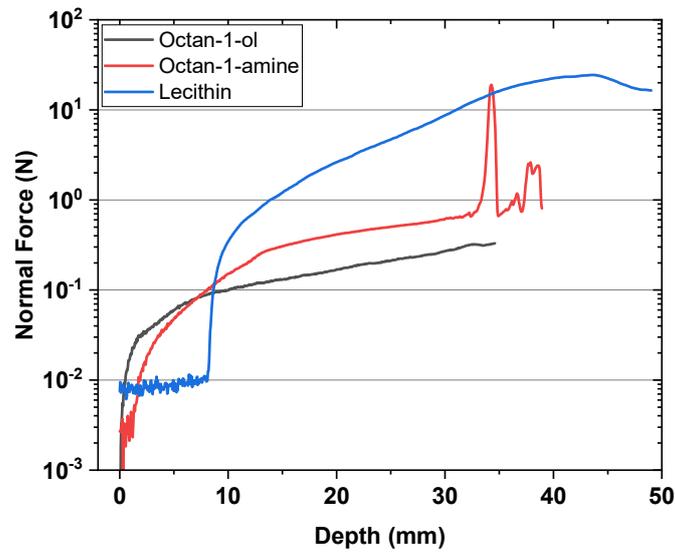

**Figure 6**: The redispersibility test after one year at rest under normal gravity. The normal force is plotted as a function of the penetration depth of the vane tool for the MRF formulations. The amounts of OM clay were kept the same but employed three different dispersant additives: octan-1-ol, octan-1-amine and lecithin.

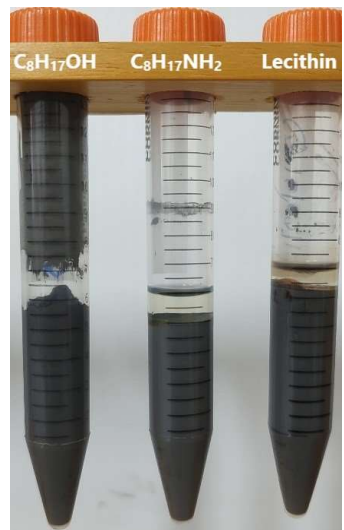

Figure 7: Picture of the test tubes containing MRF samples from the redispersibility test, after four years at rest. From left to right: octan-1-ol, octan-1-amine and lecithin.

The rheometry results for each magnetorheological fluid are summarized in Table 2:

**Table 2**: Summary of the Tau off, Tau (0.4 T), MR relative effect, G' off, Normal Force, and work results of the 3 MRF formulations.

| Test | 2.4.1 | | 2.4.3 | calculated | 2.4.6 | |
|------|-------|-----|-------|-----------|-------|-----|
| **MRF** | **$G'_{off}$ (kPa)** | **$\sigma_y$ off (Pa)** | **$\sigma_y$ 0.4 T (kPa)** | **\* Relative MR Effect** | **Normal Force (N)** | **Work (mJ)** |
| n-$C_8H_{17}OH$ | $34 \pm 1$ | $25 \pm 6$ | $19.5 \pm 3.7$ | $780 \pm 250$ | 0.33 | 3.4 |
| n-$C_8H_{17}NH_2$ | $5 \pm 1$ | $4 \pm 3$ | $19.0 \pm 2.1$ | $4750 \pm 3750$ | 19.00 | 27.6 |
| Lecithin | $1.3 \pm 0.2$ | $2 \pm 1$ | $17.0 \pm 0.8$ | $8500 \pm 3700$ | 24.40 | 414.5 |

$$* \ Relative \ MR \ effect = \frac{(\sigma_y \ on - \sigma_y \ off)}{\sigma_y \ off}$$

From Table 2, it is possible to draw some conclusions: there seems to be an inverse correlation between the storage modulus values (off-state, LVE region) and the redispersibility results: very low values of G' led to the formation of hard sediments after one year. This can be noticed for the formulations prepared with octan-1-amine and lecithin.

As the relative MR effect depends on the rheological properties in the absence of field, minimizing the G' values in off-state would be ideal. However, it is not enough to focus only on a stronger MR effect without also ensuring an easy redispersibility after sedimentation. After all, if the MRF becomes a difficult material to redisperse, it becomes useless. Note that the relative MR effect was greater for the samples with the same additives (amine and lecithin): they were the best in terms of the MR effect, but they were also the worst in terms of redispersibility.

Therefore, of the three additives tested, the only one that provided a good MR effect without compromising redispersibility was octan-1-ol.

The relative MR effect value may seem very large and it is natural for some to be skeptical of these results. Some authors opted for an approach of using an extremely high viscosity carrier fluid, such as 800 Pa.s at room temperature [30], to slow down and eliminate the old problem of sedimentation of iron particles. The drawbacks: in addition to the disadvantage of interfering with the response time under the field, this also reduces the relative MR effect, as this makes the MRF even more viscous, even in the absence of a field. Furthermore, one can imagine that it must be very difficult to disperse homogeneously and to achieve high volume fractions of ferromagnetic powder in carrier liquids with so high viscosity.

To the best of our knowledge, very few groups have studied MR fluids with *phi* values (volume fraction of particulate iron) of $\phi = 50$ vol%. A quick search on ISI Web of Science (October 11th, 2023) with the key words: "magnetorheological OR magneto-rheological OR magnetorheology OR magneto-rheology (All Fields) AND "50 vol%" OR "50 vol/vol" (All Fields), returned only 7 results. Of these articles, two were about MRE, one brought nothing useful, and one was focused on normal stress differences [31]. Volkova and Bossis [32,33] reported phi values above 50 vol%. However their works focus on shear-thickening MRF, and its applications are very different from the usual, known MR devices. Five previous works are relevant, and we quote them here, in the form of Table 3, below.

Table 3: Relative MR effect, calculated as: ($\sigma_y$ on $-\sigma_y$ off )/ $\sigma_y$ off, from works previously reported in the literature, for MRF with 50 vol% (or 90 wt.%) of iron.

| Author, year | | $\phi$ Fe (vol%) | $\sigma_y$ off (Pa) | $\sigma_y$ on (kPa) | Rel. MR effect | Note: |
|---|---|---|---|---|---|---|
| Laun et al (2008) | [31] | 46.5 | 80 | $\sim 26$ | 324 | 0.4 T |
| Böse & Ehrlich (2010) | [32] | 50.0 | $\sim 100$ | $\sim 52$ | 519 | 400 mT |
| Laun et al (2011) | | | | | | |
|    Brass plate, grooved | | * | 15 | 30.23 | 2,014 | 0.5 T |
|    Iron plate, grooved | [33] | | 16 | 40.83 | 2,550 | |
| López-López et al (2012) | | 50.0 | 120 | 0.8 | 5.67 | 26.5 kA/m |
|    Mineral oil | [34] | | | | | |
| Gómez-Ramírez et al (2012) | | 50.0 | 2 | 1 | 499 | 40 kA/m |
|    Ionic Liquid | [35] | | | | | |

* In this work, only the wt.% is declared: 90% wt/wt.

It is interesting to note, in Table 3, how the carrier fluid strongly affects the relative MR effect. López-López et al, [34] and Gómez-Ramírez et al [35], evaluated MRF with 50% vol/vol of the same carbonyl iron powder (BASF grade CC), in mineral oil (viscosity = 39.6 mPa.s @ 25 °C) and ionic liquid (viscosity = 317 mPa.s @ 25 °C). Even though pure IL is 8 fold more viscous than the mineral oil, the yield stress without a magnetic field, of the MRFs formulated with these two carriers was much lower for MRFs prepared with IL. This corroborates our results, indicating that the rheological properties of the formulation in the absence of the field (all else being equal: same volume fraction of Fe, the same type of iron powder, without magnetic field), strongly depend on the other components of the formulation.

## 4 CONCLUSIONS

Regarding the viscoelastic gels prepared with organomodified clays:

1. Among the 15 organomodified (OM) clays evaluated in this work, only three of them (the MMT clays) did not jellify when mixed with polyalphaolefin oil. Among the remaining 12 that were jellified, the rheometry results indicated clays #5, 6, and 14 (smectites, hectorites, and bentonites) were good candidates as thixotropic additives for the preparation of MR fluids. Clay #6 was chosen as the best gelling additive because it presented: the highest value of yield stress $\sigma_0 = (42 \pm 3)$ Pa; the largest storage modulus $G' = (2690 \pm 201)$ Pa; and the highest CED value = 98 mJ/m³. In addition, the complex viscosity of the clay used in the MR formulations increases with wt.%, as would be expected.

Regarding the MR fluids:

2. Using clay #6, it was possible to prepare three stable and functional MRF's by dispersing HS carbonyl iron particles in a matrix of PAO oil, all with the composition of 90 wt.% iron, 51 vol% in solids (iron + clay), and 1 vol% in clay.

3. The amplitude sweep test with the application of a magnetic field allowed us to verify the

abrupt change of the rheological properties for all fluids. The highest values of the relative MR effect were verified for fluids prepared with octan-1-amine and lecithin.

4. The three interval thixotropy tests showed that all MR fluids in this study have a good thixotropic behavior with viscosity recovery above 100% for all additives. In addition, this recovery was very fast, reaching 100% in less than 2 seconds, immediately after the shear reduction step in the third interval.

5. The redispersibility tests showed that for the MR fluids with octan-1-amine and lecithin, there was a great deterioration in the suspensions' stability after one year of preparation. This is evidenced by the high values of normal force and mechanical work recorded in the redispersibility test: $F_N$ =19.00 N and w= 27.6 mJ for the MRF with octan-1-amine and $F_N$ = 24.40 N and w= 414.5 mJ for the MRF with lecithin. On the other hand, the MRF with octan-1-ol showed excellent redispersibility results; the normal force and work values after one year were 0.33 N and 3.4 mJ, respectively.

6. The results of all tests allowed us to conclude that the MR fluid prepared with the alkanol as dispersant showed a better balance between the measured properties: $G'_{off}$, $\sigma_0$, relative MR effect, and good redispersibility after one year of preparation. The use of octan-1-ol as a dispersant and the thixotropic gel with montmorillonite #6 resulted in a functional and stable MR fluid with at least one year of shelf life.

**Acknowledgments:**

JHRR and JGFM acknowledge CAPES - *Coordenação de Aperfeiçoamento de Pessoal de Nível Superior* - for their master's scholarships granted. AJFB thanks FAPEMIG for the grants: APQ-00531-08, ETC-00044-13, ETC-00043-15, APQ-01824-17, and APQ-00620-23.

# REFERENCES

[1] Dirk H.J.A. Lukaszewicz, Carwyn Ward, Kevin D. Potter, "The engineering aspects of automated prepreg layup: History, present and future", *Composites Part B: Engineering*, vol. 43, no. 3, pp. 997-1009, 2012. DOI: 10.1016/j.compositesb.2011.12.003.

[2] A.V. Chertovich, G.V.Stepanov, E.Y. Kramarenko, and A.R. Khokhlov, (2010), "New Composite Elastomers with Giant Magnetic Response". *Macromol. Mater. Eng.*, vol. 295, no. 4, pp. 336-341, 2010. DOI 10.1002/mame.200900301.

[3] Juan de Vicente, Daniel J. Klingenberg, and Roque Hidalgo-Álvarez, "Magnetorheological fluids: a review", *Soft Matter*, vol. 7, no. 8, pp. 3701-3710, 2011. DOI: 10.1039/c0sm01221a.

[4] Hua-xia Deng, Xing-long Gong and Lian-hua Wang, "Development of an adaptive tuned vibration absorber with magnetorheological elastomer", *Smart Mater. Struct.,* vol. 15, no. 5, N111-N116, 2006. DOI: 10.1088/0964-1726/15/5/N02.

[5] Sara R. G. de Sousa, Monique P. dos Santos, Antonio J. F. Bombard, "Magnetorheological gel based on mineral oil and polystyrene-b-poly(ethene-co-butadiene)-b-polystyrene", *Smart Mater. Struct.,* vol. 28, 105016, 2019. DOI: 10.1088/1361-665X/ab3600.

[6] Modesto T. López-López, Ana Gómez-Ramírez, Juan D. G. Durán, and Fernando González-Caballero, "Preparation and Characterization of Iron-Based Magnetorheological Fluids Stabilized by Addition of Organoclay Particles", *Langmuir,* vol. 24, no. 14, pp. 7076–7084, 2008. DOI: 10.1021/la703519p.

[7] M. Ashtiani, S.H. Hashemabadi and A. Ghaffari, "A review on the magnetorheological fluid preparation and stabilization", *J. Magnetism Magn. Materials*, vol. 374, no. pp. 716-730, 2015. DOI: 10.1016/j.jmmm.2014.09.020.

[8] Yongyu Lu, He Zhou, Henan Mao, Shousheng Tang, Lei Sheng, Hu Zhang, and Jing Liu, "Liquid Metal-Based Magnetorheological Fluid with a Large Magnetocaloric Effect", *ACS Applied Materials & Interfaces*, vol. 12, no. 43, pp. 48748-48755, 2020. DOI: 10.1021/acsami.0c11153.

[9] P. P. Phulé, M. P. Mihalcin, and S. Genç, "The role of the dispersed-phase remnant magnetization on the redispersibility of magnetorheological fluids," *Journal of Materials Research*, vol. 14, no. 7, pp. 3037–3041, 1999.

[10] Makoto Kanda, Aya Kaide, Takashi Saeki, and Hiroshi Tochigi, "Preparation and rheology of magnetorheological fluid using six kinds of fumed silica as stabilizing additives", In MATEC Web of Conferences, APCChE 2019, 333, 2021, pp. 02006. DOI: 10.1051/matecconf/202133302006.

[11] Mpitloane J. Hato, Hyoung J. Choi, Hyung H. Sim, Byung O. Park, and Suprakas S. Ray, "Magnetic carbonyl iron suspension with organoclay additive and its magnetorheological properties", *Colloids and Surfaces A: Physicochem. Eng. Aspects* , vol. 377, no. 1, pp. 103–109, 2011, DOI: 10.1016/j.colsurfa.2010.12.029.

[12] F. Bergaya, B. K. G. Theng, and G. Lagaly, Eds., *Handbook of Clay Science*, Amsterdam: Elsevier, 2006.

[13] P.P. Phule; "Magnetorheological fluid"; US. Patent # 5,985,168. Nov. 16, 1999.

[14] J. R. Morillas, A J F Bombard, and Juan de Vicente; Smart Mater. Struct. 25 (2016) 015023 (10pp) doi:10.1088/0964-1726/25/1/015023.

[15] J.D. Carlson and R. Weiss; US Patent # 5,382,373; 1995.

[16] S. R. G. de Sousa, A. Leonel and A J F Bombard; Smart Mater. Struct. 29 (2020) 055039 (16pp) https://doi.org/10.1088/1361-665X/ab6abe.

[17] Podszun, Wolfgang; US Patent # 5,989,447; 1999.

[18] L.A. Poweel, W. Hu, and N.M. Wereley; "Magnetorheological fluid composites synthesized for helicopter landing gear applications"; JIMSS; 24 (9): 1043-1048 (2013); DOI: 10.1177/1045389X13476153

[19] David Chaiko, "Activation of organoclays and preparation of polyethylene nanocomposites", 2006, *e-Polymers*, 6, Available: https://www.degruyter.com/document/doi/10.1515/epoly.2006.6.1.242/html [Accessed March 28, 2023].

[20] T. G. Mezger, *The Rheology Handbook: For users of rotational and oscillatory rheometers - 4th Edition*, Hanover: Vincentz Network, 2014, pp. 146-159.

[21] T. G. Mezger, *The Rheology Handbook: For users of rotational and oscillatory rheometers - 4th Edition*, Hanover: Vincentz Network, 2014, pp. 159-174.

[22] K. Wollny, J. Läuger, and S. Huck, "Magneto Sweep – A New Method for

Characterizing the Viscoelastic Properties of Magneto-Rheological Fluids", *Applied Rheology*, vol. 12, no. 1, pp. 25-31, 2002. DOI: 10.1515/arh-2002-0003.

[23] T. G. Mezger, *The Rheology Handbook: For users of rotational and oscillatory rheometers - 4th Edition*, Hanover: Vincentz Network, 2014, pp. 72-74.

[24] S. R. G. de Sousa, and A. J. F. Bombard, "Redispersibility and its relevance in the formulation of magnetorheological fluids", in: *Magnetorheological Materials and their Applications*, S. B. Choi and W. H. Li, Eds. London: The Institution of Engineering and Technology, 2019, pp. 1-18.

[25] S. B. Ross-Murphy, *Physical Techniques for the Study of Food Biopolymers*, Dordrecht, Netherlands: Springer, 1994, pp. 362-364.

[26] T. G. Mezger, *The Rheology Handbook: For users of rotational and oscillatory rheometers - 4th Edition*, Hanover: Vincentz Network, 2014, pp. 351, 355.

[27] J. F. Steffe, *Rheological Methods in Food Process Engineering*, East Lansing MI: Freeman Press, 1996, pp. 329-332.

[28] K. Almdal, J. Dyre, S. Hvidt, and O. Kramer, "Towards a Phenomenological Definition of the Term 'Gel' ", *Polymer Gels and Networks* vol. 1, no. 1, pp. 5-17, 1993.

[29] T. F. Tadros, *Rheology of Dispersions: Principles and Applications*, Weinheim, Germany: Wiley-VCH, 2010, pp. 78-79.

[30] L. Xie, Y.T. Choi, C.R. Liao, Z. Zeng and N.M. Wereley, "Synthesis and rheological characteristics of high viscosity linear polysiloxane carrier fluid-based magnetorheological fluids"; *Smart Mater. Struct.* 31: 015041, 2022. DOI: 10.1088/1361-665X/ac3da1.

[31] H.M. Laun et al, J. Non-Newtonian Fluid Mech. 148 (2008) 47–56.

[32] Bossis, G., Ciffreo, A., Grasselli, Y. et al. Analysis of the rheology of magnetic bidisperse suspensions in the regime of discontinuous shear thickening. Rheol Acta 62, 205–223 (2023). DOI: 10.1007/s00397-023-01388-x

[33] G. Bossis, O. Volkova, Y. Grasselli and A. Ciffreo; "The Role of Volume Fraction and Additives on the Rheology of Suspensions of Micron Sized Iron Particles"; Front. Mater. 6:4. DOI: 10.3389/fmats.2019.00004

[34] H. M. Laun; C. Gabriel; Chr. Kieburg; "Twin gap magnetorheometer using ferromagnetic steel plates—Performance and validation"; J. Rheol. 54, 327–354 (2010). DOI:10.1122/1.3302804

[35] H. Boese and J. Ehrlich; "Performance of Magnetorheological Fluids in a Novel Damper with Excellent Fail-safe Behavior"; J. Intelligent Material Systems and Structures, Vol. 21—October 2010 1537; DOI: 10.1177/1045389X09351760

[36] H.M. Laun, C. Gabriel, C. Kieburg; Rheol Acta (2011) 50:141–157. DOI 10.1007/s00397-011-0531-8

[37] M.T. López-López; P. Kuzhir; J. Caballero-Hernández; L. Rodríguez-Arco; J. D. G. Duran; G. Bossis; J. Rheol. 56, 1209 (2012). DOI: 10.1122/1.4731659

[38] A. Gómez-Ramírez; M.T. López-López; F. González-Caballero; J. D. G. Durán; Rheol Acta (2012) 51:793–803. DOI 10.1007/s00397-012-0639-5.